\pdfoutput=1 
\documentclass[aps,prd,twocolumn,superscriptaddress,amsfont,graphicx,nofootinbib,preprintnumbers]{revtex4}

\usepackage{amsmath}
\usepackage{amssymb}
\usepackage{color}
\usepackage{xcolor}
\usepackage{graphicx}
\usepackage{hyperref}
\usepackage{textcomp}
\usepackage{xspace,slashed}
\usepackage[utf8]{inputenc}
\usepackage{subcaption}
\usepackage{multirow}
\usepackage{nccmath} 
\usepackage{slashed}
\usepackage[bb=boondox]{mathalfa} 

\newcommand{\ep}{\epsilon}

\newcommand{\amp}{\,\mathcal{A}}
\newcommand{\cut}{\text{Cut}}
\newcommand{\disc}{\text{Disc}}
\newcommand{\tpi}{2 \pi i}

\newcommand{\dPS}{d\text{PS}}

\begin{document}

\title{Integrated Unitarity for Scattering Amplitudes}

\author{Piotr Bargie\l{}a}
\email{piotr.bargiela@physik.uzh.ch}
\affiliation{Physik-Institut, Universit\"at Z\"urich, Winterthurerstrasse 190, 8057 Z\"urich, Switzerland}

\preprint{ZU-TH 18/24}

\begin{abstract}
We present a new method for computing multi-loop scattering amplitudes in Quantum Field Theory.
It extends the Generalized Unitarity method by constraining not only the integrand of the amplitude but also its full integrated form.
Our approach exploits the relation between cuts and discontinuities of the amplitude.
Explicitly, by the virtue of analyticity and unitarity of the S-matrix, the amplitude can be expressed in terms of lower-loop on-shell amplitudes dispersively integrated along cuts.
As both cuts and discontinuities can be computed systematically in dimensional regularization, we validated our method by reproducing the four-gluon amplitude in two-loop massless Quantum Chromodynamics.
Moreover, since our approach improves the performance of the calculation, we provide a new result for the four-loop four-point massless planar ladder Feynman integral.
It is expressed in terms of Harmonic Polylogarithms with letters 0 and 1.
\end{abstract}

\maketitle 

\allowdisplaybreaks

\tableofcontents

\section{Introduction}
\label{sec:intro}

Scattering amplitudes provide the most fundamental description of a probability amplitude for a scattering process of elementary particles in a given Quantum Field Theory.
In the small coupling region, they can be computed systematically as a perturbative expansion of multi-loop Feynman diagrams.
A fixed-order amplitude is a function of kinematics, helicity, and color.
Recently, there have been major advances in understanding the dependence of the amplitude on these three structures, see e.g. Refs~\cite{Bourjaily:2022bwx,Arkani-Hamed:2017jhn,Melia:2015ika}.
In this work, we focus on the analytic properties in the kinematic space.

Bare multi-loop scattering amplitudes are usually divergent.
Among various schemes designed to avoid such divergences, in phenomenological applications, the most popular is to perform a \textit{dimensional regularization} (dimReg) in $d=4-2\ep$ around $\ep \to 0$.
In this scheme, bare amplitude carries an additional dependence on $d$, which can be fixed to $d=4$ only after combining with appropriate ultraviolet and infrared counterterms.
There are multiple available methods of calculating scattering amplitudes in dimensional regularization (dimReg) for state-of-the-art complex processes e.g. form factor decomposition~\cite{Garland:2002ak}, generalized unitarity~\cite{Badger:2013gxa}, and numerical unitarity~\cite{Abreu:2017xsl}.
Here, we build on the unitarity arguments.

Scattering amplitude can be decomposed as a linear combination of Feynman integrals with kinematic-dependent coefficient functions.
Generalized unitarity method allows to relate these coefficients to on-shell tree-level amplitudes appearing at a unitarity cut of the original amplitude~\cite{Britto:2004nc,Anastasiou:2006jv,Ellis:2007br,Giele:2008ve,Ellis:2008ir,Badger:2008cm}.
In this work, we describe how to extend the cut decomposition beyond reconstructing coefficient functions, i.e. to also constrain the Feynman integrals, and thus the full amplitude.

Our approach exploits constraints on the kinematic dependence of Feynman integrals in dimReg arising from matching their cuts to their discontinuities~\cite{Cutkosky:1960sp,Mandelstam:1958xc,Mandelstam:1959,Mandelstam:1959bc}.
These two quantities can be computed in an algorithmic way using differential equations~\cite{Kotikov:1990kg,Henn:2013pwa,Abreu:2014cla,Abreu:2015zaa,Abreu:2021vhb} and monodromy group techniques~\cite{Bourjaily:2020wvq,Hannesdottir:2021kpd,Britto:2024mna}, respectively.
The matching can be efficiently performed either using a dispersion relation~\cite{Mandelstam:1958xc,Mandelstam:1959,Mandelstam:1959bc} or an ansatz reconstruction method.
Similar arguments have been previously successfully applied to cutting edge results using bootstrap~\cite{Dixon:2011pw}.
Our procedure builds on the developments of the early analytic S-matrix program~\cite{Cutkosky:1960sp,Mandelstam:1958xc,Mandelstam:1959,Mandelstam:1959bc,Eden:1966dnq,Remiddi:1981hn,vanNeerven:1985xr,Ball:1991bs,Kniehl:1991gu} and its modern revival~\cite{Mizera:2021fap,Mizera:2021icv,Hannesdottir:2022bmo,Mizera:2023tfe,Caron-Huot:2023vxl,Caron-Huot:2023ikn,Fevola:2023kaw,Fevola:2023fzn,Helmer:2024wax}, as well as on recent investigations in analytic properties of Feynman integrals~\cite{Abreu:2014cla,Abreu:2015zaa,Abreu:2021vhb,Bourjaily:2020wvq,Hannesdottir:2021kpd,Britto:2024mna}.

\section{Cuts}
\label{sec:cuts}

We provide here a summary of the standard procedure for calculating multi-loop multi-scale Feynman integrals based on Integration-By-Parts identities~\cite{Tkachov:1981wb,Chetyrkin:1981qh} and differential equations method~\cite{Kotikov:1990kg,Henn:2013pwa}.
With the \textit{reversed unitarity} method~\cite{Anastasiou:2002yz}, it has been applied also to cut propagators relevant for phase space integration in cross section computations.
These two approaches can be combined to deal with cut integrals with multiple external channels as required for cuts of scattering amplitudes~\cite{Abreu:2014cla,Abreu:2015zaa,Abreu:2021vhb}.

Consider an \textit{integral topology} with $L$ loop momenta $k_l$ and $N$ \textit{generalized propagators} $\mathcal{D}_i$
\begin{equation}
	\mathcal{I}_{\{n_i\}} 
	= \int \left(\prod_{l=1}^L D^d k_l \right) \, \prod_{i=1}^N \mathcal{D}_i^{-n_i} \,.
	\label{eq:inttopo}
\end{equation}
For definiteness, we choose the integration measure to be
\begin{equation}
	D^d k_l = e^{\epsilon \gamma_E} \frac{d^d k_l}{i \pi^{d/2}} \,.
\end{equation}
The set of generalized propagators consists of explicit integrand denominators as well as \textit{irreducible scalar products} (ISPs).
Each propagator can be raised to a corresponding power $n_i$.

The integrals $\mathcal{I}_{\{n_i\}}$ are in general not linearly independent.
Indeed, by exploiting the \textit{Integration-By-Parts identities} (IBPs)~\cite{Tkachov:1981wb,Chetyrkin:1981qh}
\begin{align}
	\int \left(\prod_{l=1}^L D^d k_l \right) \, \frac{\partial}{\partial k^\mu_l} \left( q^\mu \prod_{i=1}^N \mathcal{D}_i^{-n_i} \right)
	= 0
	\label{eq:IBP}
\end{align}
and the \textit{Lorentz Invariance identities}~\cite{Gehrmann:1999as}
\begin{align}
	p_{j\,\mu} \, p_{l\,\nu} \left( p_n^\nu \frac{\partial}{\partial p_{n,\mu}} - p_n^\mu \frac{\partial}{\partial p_{n,\nu}} \right)	\mathcal{I}_{\{n_i\}} = 0 \,,
	\label{eq:LI}
\end{align}
one can systematically reduce all Feynman integrals in a fixed topology to a minimal set of linearly independent \textit{Master Integrals} (MIs) $M_i$.
Following the \textit{Laporta algorithm}~\cite{Laporta:2000dsw}, this reduction has been implemented in multiple modern algorithmic tools.

Since the derivative of any MI can be decomposed into a linear combination of MIs using the IBP reduction, the MIs satisfy differential equations (DEQs)~\cite{Kotikov:1990kg}
\begin{equation}
	\partial_{x_n} M_i(\vec{x},\ep) = A_{ij}(\vec{x},\ep) \, M_j(\vec{x},\ep) \,,
	\label{eq:deqMI}
\end{equation}
where $\vec{x}$ is a set of kinematic variables in the topology~\ref{eq:inttopo}.
In Ref.~\cite{Henn:2013pwa}, it has been shown that by an appropriate choice of \textit{canonical} MIs $M^c_i$, the resulting \textit{canonical} DEQ can be solved perturbatively in $\ep$
\begin{equation}
	\partial_{x_n} M^c_i(\vec{x},\ep) = \ep \, A^c_{ij}(\vec{x}) \, M^c_j(\vec{x},\ep) \,.
	\label{eq:deqCan}
\end{equation}
The boundary conditions (BCs) to such canonical DEQ can be usually fixed by a combination of regularity constraints and direct integration.
For the example topologies in sec.~\ref{sec:examples}, see Refs~\cite{Henn:2020lye,Bargiela:2021wuy,Bargiela:2023aiy} for extended discussion.

Importantly, the same computational procedure can be applied to Feynman integrals on a cut.
The \textit{unitarity cut} of a propagator is defined via an integrand level substitution
\begin{equation}
	\frac{1}{\mathcal{D}+i\varepsilon} \to \tpi \, \delta^+(\mathcal{D}) \,,
	\label{eq:cutprop}
\end{equation}
where $\delta^+(q^2) = \delta(q^2) \, \theta(q_0)$.
A \textit{cut integral topology} $\mathcal{I}_{\{n_i\};\{c_j\}}$ results from cutting $C$ propagators belonging to a set of indices $c_j \in \{1,...,N\}$ of the original integral topology.
Note that before explicitly evaluating the cut boundary MIs, the set $\{c_j\}$ serves only as a criterion on the property of the corresponding propagators.
Indeed, any cut integral not supported on the full cut $\{c_j\}$ vanishes.
This reduces the number of subsectors by a factor of $2^C$, thus decreasing the complexity of the corresponding IBP reduction, as well as the number of MIs.
On the other hand, when directly integrating, it is important to choose a basis of MIs without any higher denominator powers for the cut definition~\ref{eq:cutprop} to apply.
In practice, this basis can be usually found by IBP-relating integrals with higher denominator powers to those with more ISPs~\cite{Sogaard:2014ila,Zeng:2017ipr}.
The definition of a cut in an arbitrary channel $s$ reads
\\
\scalebox{0.85}{\parbox{1.0\linewidth}{
		\begin{equation}
			\begin{split}
				&\cut_{s} \mathcal{I}_{\{n_i\}} = \sum_{\{c_j\} \in \mathcal{C}_{s}} \mathcal{I}_{\{n_i\};\{c_j\}} \,, \\
				&= \sum_{\{c_j\} \in \mathcal{C}_{s}}
				\int \left( \prod_{l=1}^L D^d k_l \right) \, 
				\left( \prod_{i \notin \{c_j\}} \mathcal{D}_i^{-n_i} \right) \,
				\prod_{m \in \{c_j\}} \delta_{1,n_m} \tpi \, \delta^+(\mathcal{D}_m) \,, 
			\end{split}
		\label{eq:cutdef}
		\end{equation}
}}
\\
where $n_i \leq 1$, and $\mathcal{C}_{s}$ denotes all the sets of propagators which cut in the $s$ channel the diagrammatic representation of the integral $\mathcal{I}_{\{n_i\}}$ into two lower-loop diagrams, i.e. the propagators along each $s=p^2$ cut whose momenta sum to $p^\mu$.
The cut BC integrals of the canonical DEQ can be computed using e.g.  ellin-Bernes parametrization~\cite{Dubovyk:2022obc}.

\section{Discontinuities}
\label{sec:disc}

\begin{figure}[h]
	\centering
	\includegraphics[width=0.25\textwidth]{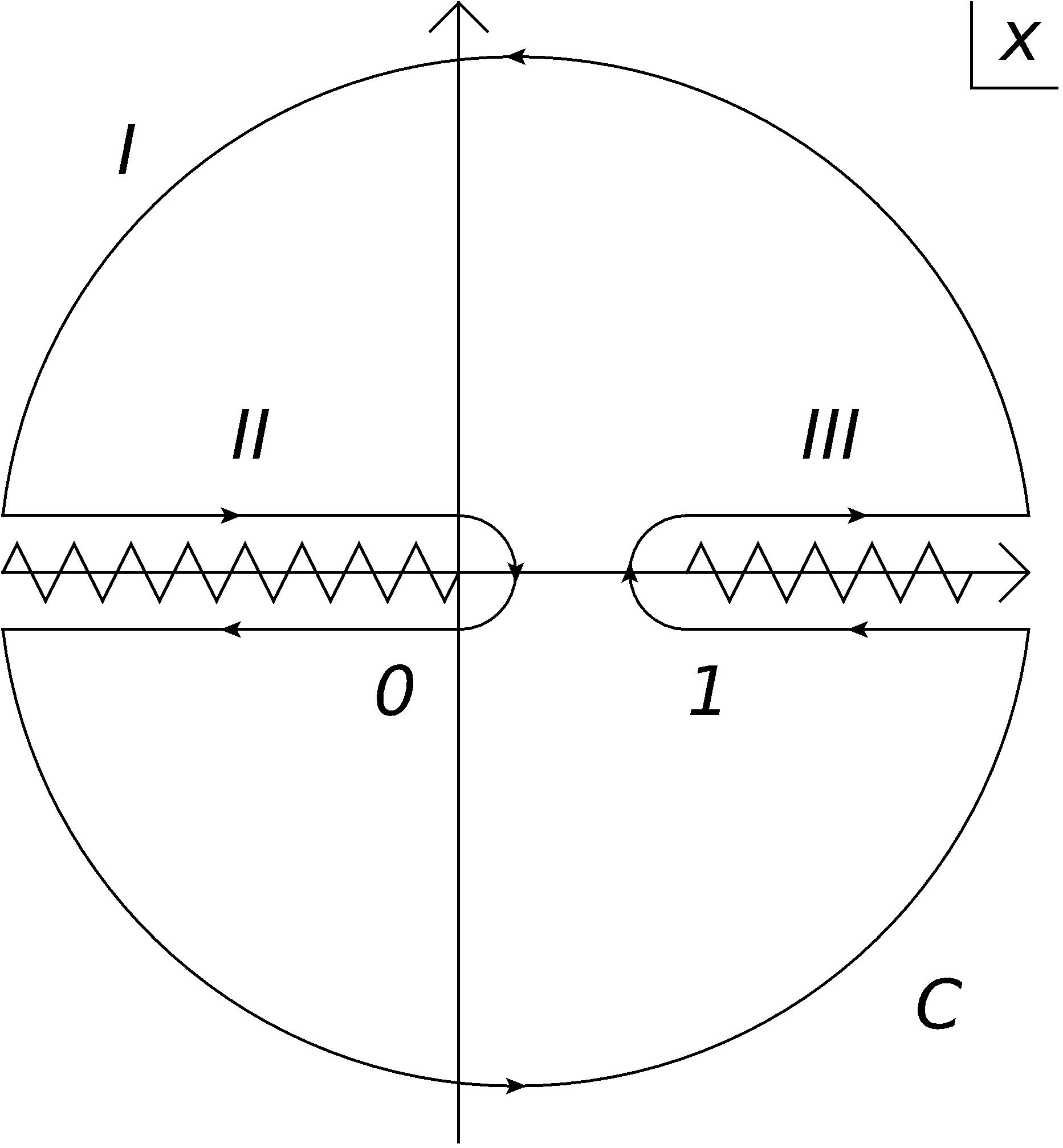}
	\caption{Analytic structure of the four-particle massless scattering~\ref{eq:proc} and the dispersive integration contour~\ref{eq:disp}.}
	\label{fig:contour}
\end{figure}

Ref.~\cite{Bourjaily:2020wvq} introduces a method based on monodromy group properties for computing sequential discontinuities of iterated integrals.
Here, we provide a summary of results relevant for the further discussion~\footnote{Note slight differences in signs between our notation and Ref.~\cite{Bourjaily:2020wvq}.}.

From now on, let us focus on a four-particle massless scattering process
\begin{equation}
	i_1(p_1) + i_2(p_2)  \to i_3(-p_3) + i_4(-p_4) \,.
	\label{eq:proc}
\end{equation}
Its kinematics is fully parametrized by the three Mandelstam variables
\begin{equation}
	s = (p_1+p_2)^2 \,, \quad
	u = (p_2+p_3)^2 \,, \quad
	t = (p_1+p_3)^2 \,,
\end{equation}
which are related by a momentum conservation relation
\begin{equation}
	s+t+u = 0 \,.
\end{equation}
It is convenient to define a dimensionless ratio
\begin{equation}
	x = -\frac{t}{s}
	\label{eq:x}
\end{equation}
which carries all the nontrivial kinematic dependence.
The physical region is defined through $s>0$, $t<0$, and $u<0$, thus the resulting dimensionless ratio is positive $0<x<1$.

Let us describe the analytic properties of the scattering amplitude $\amp$ in these variables.
There are poles corresponding to single-particle production at $s=0$, $t=0$, and $u=0$, which in variable $x$ translates into poles at $x_p \in \{0,1,\infty\}$.
In addition, there are branch cuts corresponding to multi-particle production along $s>0$, $t>0$, and $u>0$, which translates into $x<0$ and $x>1$.
It has been show that these poles and branch cuts are the only singularities appearing up to three-loop order~\cite{Henn:2013pwa,Henn:2020lye,Bargiela:2021wuy}.
The analytic structure on the complex $x$ plane is depicted in Fig.~\ref{fig:contour}.

Consider a Harmonic Polylogarithm (HPL) $G(\vec{\alpha},x)$ of transcendental weight $n$ and letters $\alpha_k \in \{0,1\}$~\cite{Remiddi:1999ew,Gehrmann:2001pz}.
Such HPLs span a complete transcendental basis relevant for our examples in sec.~\ref{sec:examples}.
We seek a systematic construction of the discontinuity $\disc_{x_p} \, G(\vec{\alpha},x)$ around $x_p \in \{0,1,\infty\}$.
We start by defining a \textit{vector of derivatives} for $i \in \{0,...,n\}$ via
\begin{equation}
	\mathcal{V}_i =
	\begin{cases}
		1 & \text{if } i=0 \,, \\
		(-1)^{\text{\# of 1s}} \, G(\alpha_{n+1-i},...,\alpha_n,x) & \text{if } i>0 \,.
	\end{cases}
\end{equation}
The differential equation $d \, \mathcal{V} = \mathcal{V} \cdot \omega$ is defined by the \textit{connection matrix}
\begin{equation}
	\omega_{ij} = \frac{dx}{x-\alpha_{n-i}} \, \delta_{i+1,j} \,,
\end{equation}
where $i,j \in \{0,...,n\}$.
The $(n+1)$ independent solutions to this differential equation can be collected in the \textit{variation matrix}
\begin{equation}
	\mathcal{M}_\gamma = \mathcal{P} \, e^{\int_\gamma \omega} \,.
\end{equation}
General solution arises from integrating along a path $\gamma$ which leads to a generic point $x$
\begin{equation}
	\left( \mathcal{M}_{\to x} \right)_{ij} = \sum_{k=0}^{n} (-1)^{\text{\# of 1s}} \, G(\alpha_{n-i},...,\alpha_{n-i-k+1},x) \, \delta_{i+k,j} \,,
\end{equation}
where $G(x)=1$.
By choosing appropriate integration paths $\gamma$, we define \textit{monodromy matrices} through
\begin{equation}
	\begin{split}
		\mathcal{M}_0 &= \mathcal{M}_{\circlearrowleft_0} \,, \\
		\mathcal{M}_1 &= \mathcal{M}_{\to 1} \, \mathcal{M}_{\circlearrowright_1} \, \mathcal{M}_{\to 1}^{-1} \,.
	\end{split}
\end{equation}
We do not provide explicit expressions in terms of HPLs any more as they become more complicated.
Note, that the $\tpi$ factor arises from the closed integration path $\circlearrowleft$, while $\zeta$ values stem from the evaluation at $x=1$.
Finally, we define \textit{discontinuities} around points $x=\{0,1,\infty\}$ as the $(0,n)$ element each of the resulting matrices
\begin{equation}
	\begin{split}
		\disc_0 &= (\mathbb{1}-\mathcal{M}_0) \cdot \mathcal{M}_{\to x} \,, \\
		\disc_1 &= -(\mathbb{1}-\mathcal{M}_1) \cdot \mathcal{M}_{\to x} \,, \\
		\disc_\infty &= (\mathbb{1} - \mathcal{M}_0 \cdot \mathcal{M}_1) \cdot \mathcal{M}_{\to x} \,.
	\end{split}
\end{equation}
Importantly, the discontinuity in a fixed channel is related the unitarity cut~\ref{eq:cutdef} in this channel
\begin{equation}
	\disc_0 = \cut_{t} \,,\quad
	\disc_1 = \cut_{u} \,,\quad
	\disc_\infty = \cut_{s}
	\label{eq:disccut}
\end{equation}
computed as described in sec.~\ref{sec:cuts}.

Note that Feynman integrals in dimReg can have more discontinuities in the same channel than their diagrammatic cuts.
For example, a one-loop one-off-shell massless triangle integral has a double discontinuity at order $\ep^0$, while only a single cut.
In general, at weight $w$, applying $w$ sequential discontinuities introduced in Ref.~\cite{Bourjaily:2020wvq} in an order reverse to $\vec{\alpha}$ uniquely picks up a coefficient of an HPL $G(\vec{\alpha},x)$.
However, only up to an $L$-fold discontinuity can be directly related to cuts, see discussion in Refs~\cite{Britto:2024mna,Abreu:2014cla,Abreu:2015zaa}.

\section{Dispersion relation}
\label{sec:disp}

In this section, we connect the discussion of cuts in sec.~\ref{sec:cuts} and discontinuities in sec.~\ref{sec:disc} to the notion of the dispersion relation~\cite{Mandelstam:1958xc,Mandelstam:1959,Mandelstam:1959bc,Eden:1966dnq,Remiddi:1981hn,vanNeerven:1985xr,Ball:1991bs,Kniehl:1991gu,Abreu:2014cla,Mizera:2023tfe}.
Since both cuts and discontinuities can be computed algorithmically, the constraints arising from matching them can also be systematically applied.

Let us assume analyticity of the scattering amplitude $\amp(x)$ for the process~\ref{eq:proc} in the whole complex $x$ plane except for the two branch cuts depicted in Fig.~\ref{fig:contour}.
Consider Cauchy's integral formula
\begin{equation}
	\amp(z) = \frac{1}{\tpi} \oint_C \frac{\amp(x) dx}{x-z}
	\label{eq:disp}
\end{equation}
for an integration contour $C$ in Fig.~\ref{fig:contour}.
In the following, we will omit the $s$ dependence of the amplitude $\amp$ since an overall scale dependence can be reconstructed separately from dimension analysis argument.
Let us split the integration contour $C$ into three pieces, the arc at infinity (I), the keyhole around $x<0$ (II), and the keyhole around $x>1$ (III).
The integral along (I) is a constant, along (II) is equal to the discontinuity across the $x<0$ branch cut, and along (III) is equal to the discontinuity across the $x>1$ branch cut
\begin{equation}
	\amp(z) = c_\infty + \frac{1}{\tpi}
	\left( \int_0^\infty \disc_0 + \int_1^\infty \disc_1 \right)
	\frac{\amp(x) dx}{x-z} \,.
\end{equation}
Since $c_\infty$ cancels in the difference $\amp(z) - \amp(z_0)$ for any fixed value $\amp_0 = \amp(z_0)$, we can write
\begin{widetext}	
	\begin{equation}
		\amp(z) = \amp_0+ \frac{1}{\tpi}
		\left( \int_0^\infty \disc_0 + \int_1^\infty \disc_1 \right)
		\left( \frac{1}{x-z} - \frac{1}{x-z_0} \right) \amp(x) dx \,.
	\end{equation}
	By the virtue of unitarity, we can use the relation between discontinuities and cuts~\ref{eq:disccut} and obtain
	\begin{equation}
		\amp(z) = \amp_0 + \frac{1}{\tpi} 
		\left( \int_0^\infty \cut_{t} + \int_1^\infty \cut_{u} \right)
		\left( \frac{1}{x-z} - \frac{1}{x-z_0} \right) \amp(x) dx
	\end{equation}
	Finally, $t$ and $u$ cut of the scattering amplitude $\amp(x)$ can be expressed in terms of a sum over cuts~\ref{eq:cutdef} of two lower-loop on-shell amplitudes $\amp_{L}$ and $\amp_{R}^*$ integrated over the phase space along the corresponding set of indices $\{c_i\}$.
	We will refer to the resulting dispersion relation as the \textit{Integrated Unitarity equation}
	\begin{equation}
		\begin{split}
			\amp(z) = \amp_0 + \frac{1}{\tpi}&
			\bigg(\int_0^\infty \sum_{\{c_i\} \in \mathcal{C}_{t}} \int \dPS_{t,\{c_i\}} \amp_{t,\{c_i\},L} \amp_{t,\{c_i\},R}^* \\
			&+ \int_1^\infty \sum_{\{c_j\} \in \mathcal{C}_{u}} \int \dPS_{u,\{c_j\}} \amp_{u,\{c_j\},L} \amp_{u,\{c_j\},R}^* \bigg)
			\left( \frac{1}{x-z} - \frac{1}{x-z_0} \right) dx \,.
		\end{split}
		\label{eq:IUeq}
	\end{equation}
	\begin{figure}[h]
		\centering
		\includegraphics[width=0.8\textwidth]{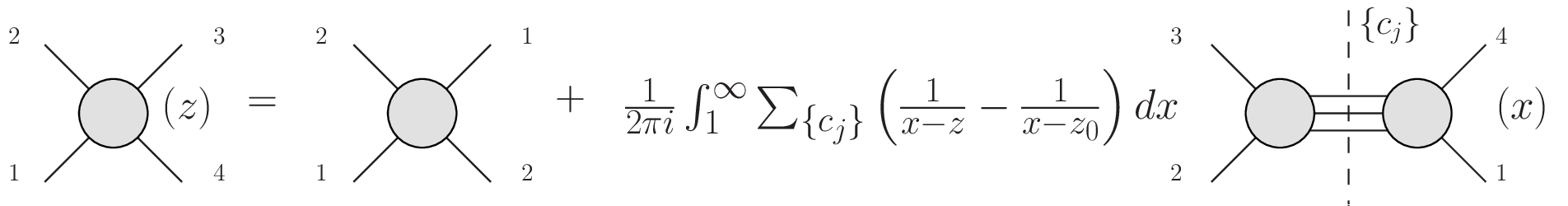}
		\caption{Planar diagrammatic Integrated Unitarity equation~\ref{eq:IUeq} for $z_0=0$.}
		\label{fig:IU}
	\end{figure}	
\end{widetext}
We depict this factorization phenomenon diagrammatically in Fig.~\ref{fig:IU} for $z_0=0$ and assuming planarity in the $s-u$ channel.
Note that, at $z_0=0$, the kinematics simplifies to $p_4=-p_2$, $p_3=-p_1$, and thus $u=-s$.
At a loop order $L$, for the kinematics~\ref{eq:proc}, the number of propagators in an underlying integral topology~\ref{eq:inttopo} drops by $L$, which reduces the number of sectors in the IBP reduction~\ref{eq:IBP} by a factor of $2^L$.

Note, that the Integrated Unitarity equation~\ref{eq:IUeq} extends the Generalized Unitarity~\cite{Britto:2004nc,Anastasiou:2006jv,Ellis:2007br,Giele:2008ve,Ellis:2008ir,Badger:2008cm,Badger:2013gxa,Abreu:2017xsl} method from integrand level to the integral level of the amplitude, i.e. it allows to reconstruct both Feynman integrals and their coefficients from cuts.
Indeed, the above derivation is true for any function $\amp(z)$ as long as it has an analytic structure as in Fig.~\ref{fig:contour} and  its cuts have a physical interpretation~\ref{eq:cutdef}.
In particular, this discussion also holds for Master Integrals.
We will now describe three separate methods of calculating the amplitude using the Integrated Unitarity equation~\ref{eq:IUeq}.

\paragraph{Explicit integration}
\label{sec:explicitInt}

The most direct way is to perform the dispersive integrals in eq.~\ref{eq:IUeq} explicitly.
These integrals are convergent for pure weight integrands, e.g. cut canonical MIs.
However, for the full amplitudes, they may diverge at some integration boundaries due to the presence of rational functions in the integrand.
Is such case, the procedure leading to eq.~\ref{eq:IUeq} has to be repeated with $\amp(z) \, S(z)$ substituted instead of $\amp(z)$, where $S(z)$ represents the so called \textit{subtraction term}~\cite{Mandelstam:1958xc,Mandelstam:1959,Mandelstam:1959bc}.
For the examples considered in sec.~\ref{sec:examples}, we have $S(x) = \frac{(1-x)^p x^q}{(x-z_1)^r}$, for some regular point $z_1$.
As a result, there is a new term $-\underset{x \to z_1}{\text{Res}} \amp(x) \, S(x) \left( \frac{1}{x-z} - \frac{1}{x-z_0} \right)$ at the right-hand side of the corresponding eq.~\ref{eq:IUeq}.
Since for higher-poles, the residue involves derivatives of the amplitude, the direct integration is not the most efficient approach.
We resolve this issue in the following two methods.

\paragraph{Ansatz reconstruction using discontinuities and an evaluation at fixed points}
\label{sec:ansatz2d}

The constraints captured by the dispersion relation~\ref{eq:IUeq} are equivalent to requiring
\begin{equation}
\begin{cases}
	\disc_0 \amp &= \cut_{t} \amp \\
	\disc_1 \amp &= \cut_{u} \amp \\
	\amp(z_i) &= \amp_i \,,
\end{cases}
\label{eq:constraints}
\end{equation}
where the set of regular points $z_i$ is required to reconstruct also the purely rational terms in the amplitude.
Therefore, the amplitude can be reconstructed by matching the left-hand side constraints applied on the ansatz
\begin{equation}
	\amp(z) = \ep^{\#}\sum_{n \geq 0} \ep^n \sum_{\vec{\alpha} \, : \, |\vec{\alpha}| \leq n} r_{n,\vec{\alpha}}(z) \,  G(\vec{\alpha},z) \,.
	\label{eq:ansatz}
\end{equation}
with the precomputed right-hand side expressions.
We denote by $\ep^{\#}$ a process-specific overall factor, while by $r_{n,\vec{\alpha}}(z)$ the unknown coefficients consisting of rational functions of $z$ and transcendental numbers of weight not exceeding $n-|\vec{\alpha}|$, where $|\vec{\alpha}|$ is a transcendental weight of an HPL $G(\vec{\alpha},z)$.
Note that for the canonical MIs, the form of the ansatz simplifies such that $r_{n,\vec{\alpha}}(z)=r_{n,\vec{\alpha}}$ are constants of transcendental weight equal to $n-|\vec{\alpha}|$.
It is also worth pointing out that in order to ensure the completeness of the ansatz, one needs to combine the letters $\vec{\alpha}$ resulting from the cut amplitudes with the ones corresponding to the discontinuities.
For our purposes, the cut MIs generate only the letters $\alpha_i \in \{0,1\}$, which coincide with the discontinuity letters.

\paragraph{Ansatz reconstruction using only discontinuities}
\label{sec:ansatz3d}

The amplitude $\amp(z)$ is also uniquely fixed by requiring $\disc_\infty \amp = \cut_{s} \amp$ instead of $\amp(z_i) = \amp_i$ in eq.~\ref{eq:constraints}.
It is because the crossed amplitude $\amp(s \leftrightarrow t) = \amp(1/x)$ is constrained by $\disc_\infty \amp(x)$.
This provides an alternative set of constraints that allow to reconstruct the ansatz~\ref{eq:ansatz}.
In sec.~\ref{sec:examples}, we will see that sometimes one of these sets provides a more efficient method than the other.
Note, that explicitly computing the discontinuities of HPLs as described in sec.~\ref{sec:disc} is required in the two ansatz reconstruction methods~\ref{sec:ansatz2d} and~\ref{sec:ansatz3d} but not in the direct integration approach~\ref{sec:explicitInt}.

Note that in any of the three above methods, the constraints originating in only the two discontinuities $\disc_0$ and $\disc_1$ fix the all the coefficients of transcendental functions in the amplitude $\amp(z)$.
With the ansatz fitting methods~\ref{sec:ansatz2d} and~\ref{sec:ansatz3d}, the remainder is a rational function, while with the explicit integration method~\ref{sec:explicitInt}, it is a constant.

\section{Examples}
\label{sec:examples}

\subsection{One-loop Master Integrals}
\label{sec:1LMIs}

\begin{figure}[h]
	\centering
	\includegraphics[width=0.35\textwidth]{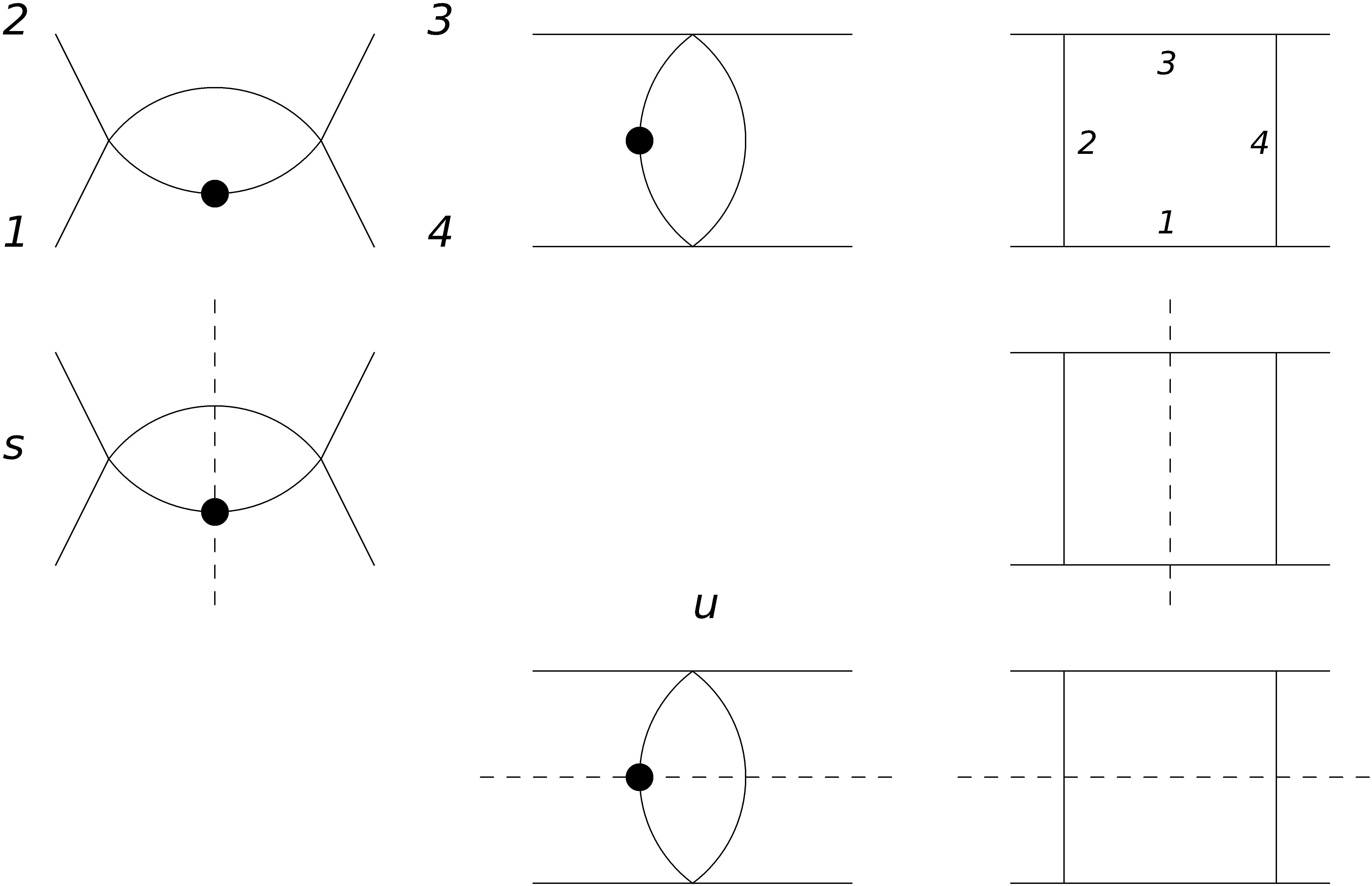}
	\caption{One-loop Master Integrals, their $s$ and $u$ cuts, and the propagator labels in the top sector.}
	\label{fig:1LMIs}
\end{figure}

Consider one-loop MIs for our four-point massless kinematics~\ref{eq:proc}.
The integral topology~\ref{eq:inttopo} is defined by $L=1$ loops and $N=4$ propagators
\begin{equation}
	\mathcal D_i \in \{(k_1)^2, (k_1+p_1)^2, (k_1+p_{12})^2, (k_1+p_{123})^2\} \,,
	\label{eq:1Ltopo}
\end{equation}
where we abbreviate $p_{ij}=p_i+p_j$ and $p_{ijk}=p_i+p_j+p_k$.
The canonical MIs read
\begin{equation}
	M^c_i \in \{\ep(2\ep-1) \, \mathcal{I}_{1,0,1,0}, \, 
	\ep(2\ep-1) \, \mathcal{I}_{0,1,0,1}, \,
	\ep^2(x-1) \, \mathcal{I}_{1,1,1,1}\} \,,
\end{equation}
see Fig.~\ref{fig:1LMIs}.
Due to the simplicity of our one-loop MIs, we will show how to obtain them using three separate methods introduced in sec.~\ref{sec:disp}.
To this end, let us compute the following set of redundant quantities $M^c(x=0,\ep)$, $\cut_t M^c$, $\cut_u M^c$, and $\cut_s M^c$.

Note that $\cut_t M^c=0$ due to the planarity in the $s-u$ channel.
For the remaining $s$ and $u$ cuts, the number of MIs supported on these cuts drops from 3 to 2
\begin{equation}
\begin{split}
	\cut_s M^c_i &\in \{\ep(2\ep-1) \, \mathcal{I}_{1,0,1,0;1,3}, \,\, 
	0, \,\,
	\ep^2(x-1) \, \mathcal{I}_{1,1,1,1;1,3}\} \,, \\
	\cut_u M^c_i &\in \{0, \,\,
	\ep(2\ep-1) \, \mathcal{I}_{0,1,0,1;2,4}, \,
	\ep^2(x-1) \, \mathcal{I}_{1,1,1,1;2;4}\} \,,
\end{split}
\end{equation}
see Fig.~\ref{fig:1LMIs}.
This reduction is also reflected in the form of the canonical DEQ~\ref{eq:deqCan} with the matrix
\begin{equation}
A^c(x) = 
\left(
\begin{array}{ccc}
	{\color{blue} \mathbf{0}} & {\color{violet} \mathbf{0}} & {\color{blue} \mathbf{0}} \\
	{\color{violet} \mathbf{0}} & {\color{red} \mathbf{\frac{1}{1-x}}} & {\color{red} \mathbf{0}} \\
	{\color{blue} \mathbf{\frac{2}{x}}} & {\color{red} \mathbf{\frac{2}{x}+\frac{2}{1-x}}} & \frac{1}{x}+\frac{1}{1-x} \\
\end{array}
\right) \,,
\end{equation}
where we coloured the rows and columns vanishing either on the $s$ {\color{red} \textbf{cut}} or the $u$ {\color{blue} \textbf{cut}}.
In addition, the corresponding BCs to each of the cut DEQs can be fixed by computing only a single cut bubble integral.
As a result, we obtain e.g. $\cut_u M^c_3 = \ep \tpi (2 + \ep \tpi + 2 \ep^2 \left( G(0,1,x) - G(1,1,x) - \frac{9}{2} \zeta_2 \right) + \mathcal{O}(\ep^4))$.
In addition, at $x=0$, the kinematics simplifies, and thus the integral topology reduces to $\mathcal D_{0,i} \in \{(k_1)^2, (k_1+p_1)^2, (k_1+p_{12})^2\}$.
The value $M^c_0(\ep)$ can be evaluation e.g. with \texttt{Mathematica} package \texttt{AMFlow}~\cite{Liu:2022chg}.

Following sec.~\ref{sec:disp}, the MIs $M^c$ can be computed in three ways i.e. by an explicit integration~\ref{sec:explicitInt} using $M^c(x=0,\ep)$, $\cut_t M^c$, and $\cut_u M^c$, by matching~\ref{sec:ansatz2d} $M^c(x=0,\ep)$, $\cut_t M^c$, and $\cut_u M^c$ with the constraints $M^c_0$, $\disc_0 M^c$, and $\disc_1 M^c$ of an ansatz~\ref{eq:ansatz}, or by matching~\ref{sec:ansatz3d} $\cut_s M^c$, $\cut_t M^c$, and $\cut_u M^c$ with the constraints $\disc_\infty M^c$, $\disc_0 M^c$, and $\disc_1 M^c$ of an ansatz~\ref{eq:ansatz}.
As discussed in sec.~\ref{sec:disp}, using only $\cut_t M^c$ and $\cut_u M^c$ in any of the three above methods fixes all the HPL coefficients except for a constant remainder.
Using all three methods separately, we found agreement with the well-known one-loop results to higher order in $\ep$.

\subsection{Higher-loop integral and amplitude checks}
\label{sec:checks}

\begin{figure}[h]
	\centering
	\includegraphics[width=0.35\textwidth]{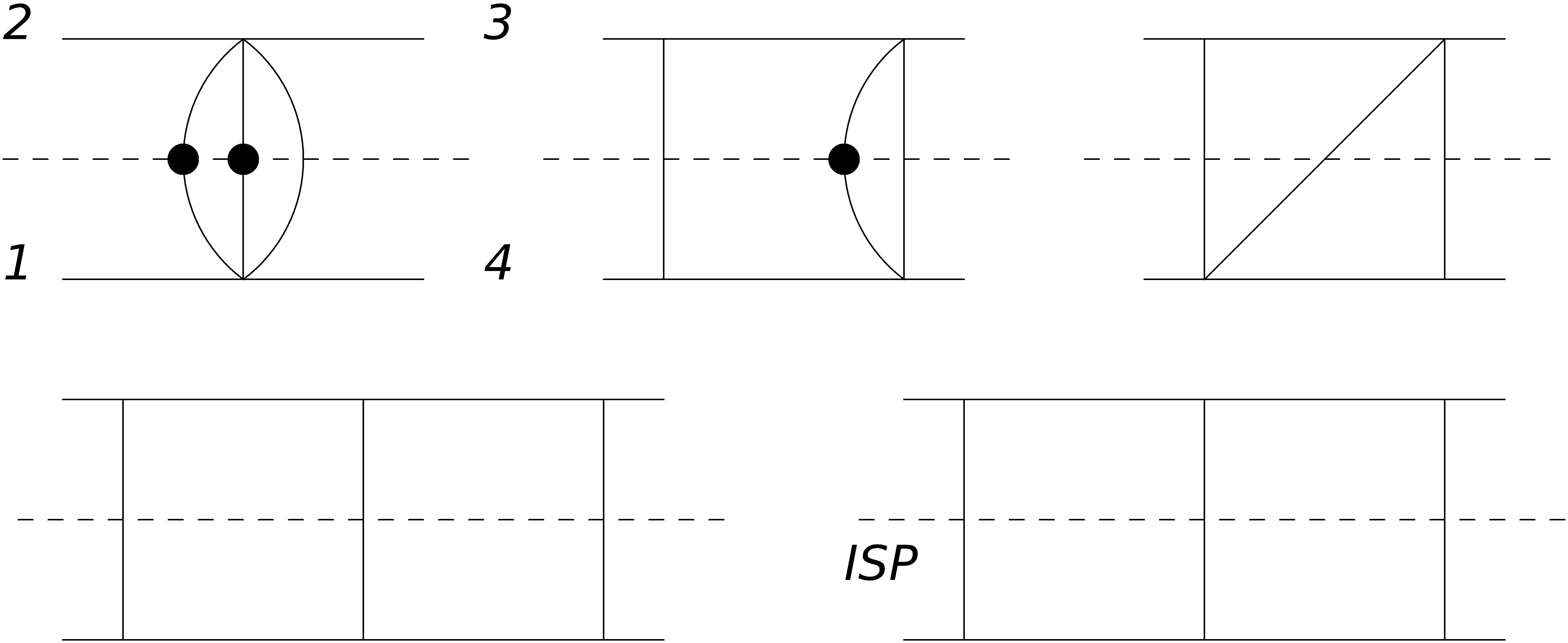}
	\caption{Two-loop planar Master Integrals cut in the $u$ channel.}
	\label{fig:2LMIs}
\end{figure}

We have performed multiple checks of the methods introduced in sec.~\ref{sec:disp} for both Master Integrals and amplitudes for the process~\ref{eq:proc} to higher loop orders.

In term of the MIs, we have recomputed all the two-loop planar, two-loop nonplanar, and three-loop planar ladder topologies.
Due to the presence of higher-order cuts in the planar ladder topologies, the methods \ref{sec:explicitInt} and \ref{sec:ansatz2d} are more efficient than \ref{sec:ansatz3d}.
Contrarily, all three methods have similar complexity for the two-loop nonplanar topology since it has 1 nonvanishing cut in each of the 3 Mandelstam channels.
Note that for this topology, $\amp_0$ cannot be evaluated at $z_0=0$ in the methods \ref{sec:explicitInt} and \ref{sec:ansatz2d} since is no longer a regular point.
Let us also compare the reduction in the computational complexity for ladder MIs planar in the $s-u$ channel with $L$ boxes in the $s$ channel at $L=2$ and $L=3$ loops.
Since $\cut_t M^c=0$, such MIs can be reconstructed from a single cut in the $u$ channel.
The number of MIs on this cut reduces from 8 to 5 at two loops, see Fig.~\ref{fig:2LMIs}, and from 26 to 17 at three loops.
Note, that cutting the corresponding propagators decreases the number of subsectors in the IBP reduction necessary to derive the DEQ by a factor of 8 at two loops and 16 at three loops.
Importantly, using regularity arguments, all the BCs of the corresponding cut DEQs can be related to 1 BC at two loops, and 2 at three loops.
For all considered results, we have found agreement with the literature~\cite{Smirnov:2003vi,Henn:2013pwa,Henn:2020lye,Bargiela:2021wuy} up to weight 8.

Regarding checks of the amplitudes, we have focused on the bare single-color-trace contribution to the four-gluon scattering at one and two loops in massless Quantum Chromodynamics (QCD).
Following the form factor based method of computing scattering amplitudes~\cite{Garland:2002ak}, we used \texttt{qgraf}~\cite{Nogueira:1991ex} to generate Feynman diagrams, \texttt{FORM}~\cite{Vermaseren:2000nd} to perform spinor algebra, and \texttt{reduze}~\cite{Studerus:2009ye,vonManteuffel:2012np} for the IBP reduction.
As explained in sec.~\ref{sec:disp}, we used method~\ref{sec:explicitInt} only as a self-consistency check of the of the Integrated Unitarity equation~\ref{eq:IUeq}.
The efficiency comparison of the other two methods~\ref{sec:ansatz2d} and~\ref{sec:ansatz3d} is the same as for the MI discussion above.
The cut structure for planar and nonplanar amplitude contributions is also the same as for their corresponding MIs.
Due to the presence of integrand numerators appearing in gauge theory, the required IBP reduction has to be performed with higher powers of ISPs in comparison to MI derivation.
In comparison to the full uncut IBP reduction, the number of integrals to IBP reduce drops by a factor of 2 at one loop, 8 at planar two loops, and 4 at nonplanar two loops per cut.
It is because we neglect the integrals not supported on the cuts e.g. scaleless integrals.
Another difference with respect to the MI computation is requiring the constraint $\amp(z_i) = \amp_i$ at multiple points $z_i$ instead of just one for MIs.
As explained in~\ref{sec:ansatz2d}, they allow for the reconstruction of the rational terms, instead of only a constant like for the MIs.
Using both methods ~\ref{sec:ansatz2d} and~\ref{sec:ansatz3d}, we reconstructed the considered amplitudes to higher orders in $\ep$ for all external helicity states.

\subsection{Four-loop planar ladder integral}
\label{sec:4LMIs}

\begin{figure}[h]
	\centering
	\includegraphics[width=0.3\textwidth]{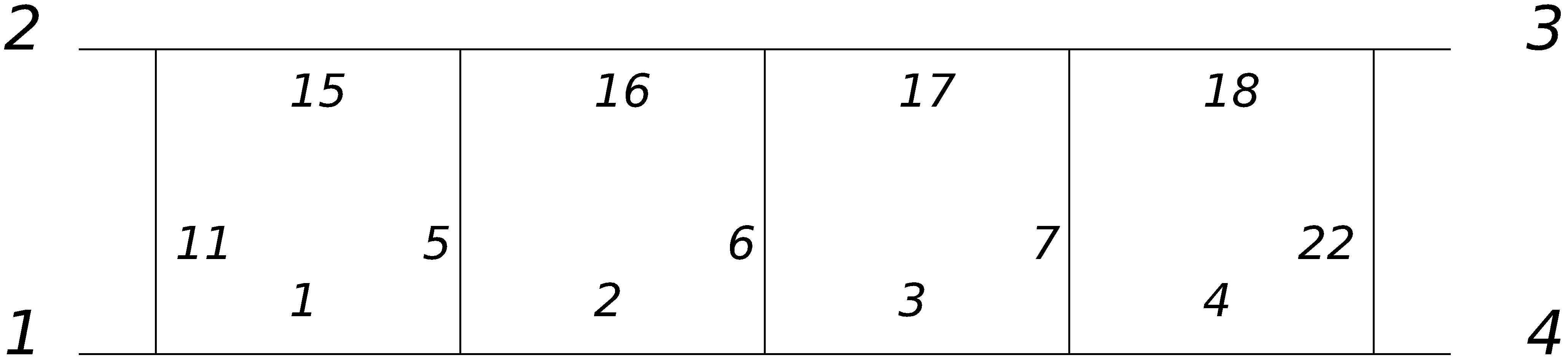}
	\caption{Four-loop planar ladder integral together with the propagator labels.}
	\label{fig:4LMIs}
\end{figure}

Consider four-loop planar ladder integral for the kinematics~\ref{eq:proc}, see Fig.~\ref{fig:4LMIs}.
It is one of the top sector integral topologies for a generic four-point massless scattering amplitude.
This integral topology~\ref{eq:inttopo} is defined by $L=4$ loops and $N=22$ generalized propagators
\begin{equation}
	\begin{split}
		\mathcal D_i \in \{&(k_1)^2, (k_2)^2, (k_3)^2, (k_4)^2, \\
		&(k_{12})^2, (k_{23})^2, (k_{34})^2, (k_{14})^2, (k_{13})^2, (k_{24})^2, \\
		&(k_1+p_1)^2, (k_2+p_1)^2, (k_3+p_1)^2, (k_4+p_1)^2, \\
		&(k_1+p_{12})^2, (k_2+p_{12})^2, (k_3+p_{12})^2, (k_4+p_{12})^2, \\
		&(k_1-p_{4})^2, (k_2-p_{4})^2, (k_3-p_{4})^2, (k_4-p_{4})^2\} \,,
	\end{split}
\end{equation}
with ISPs for $i \in \{8,9,10,12,13,14,19,20,21\}$.
In this topology, there are no cuts in the $t$ channel, a single cut in the $u$ channel, and multiple cuts in the $s$ channel.
In order to avoid computing multiple cuts, it is efficient to reconstruct the ladder integral $M_{\text{lad}}$ with the method~\ref{sec:ansatz2d} effectively from $\disc_1 M_{\text{lad}}$ and $M_{\text{lad}}(x=0,\ep)$.
Therefore, we are interested in deriving and solving a canonical DEQ in the $u$ cut channel $\{c_j\}=\{5,6,7,11,22\}$.
Since the leading singularity of the original uncut ladder integral is proportional to $\ep^{-8}$, we seek a solution on the cut to trascendental weight 7 in order to reconstruct the finite $\ep^{0}$ term of the uncut ladder.

\begin{figure}[h]
	\centering
	\includegraphics[width=0.45\textwidth]{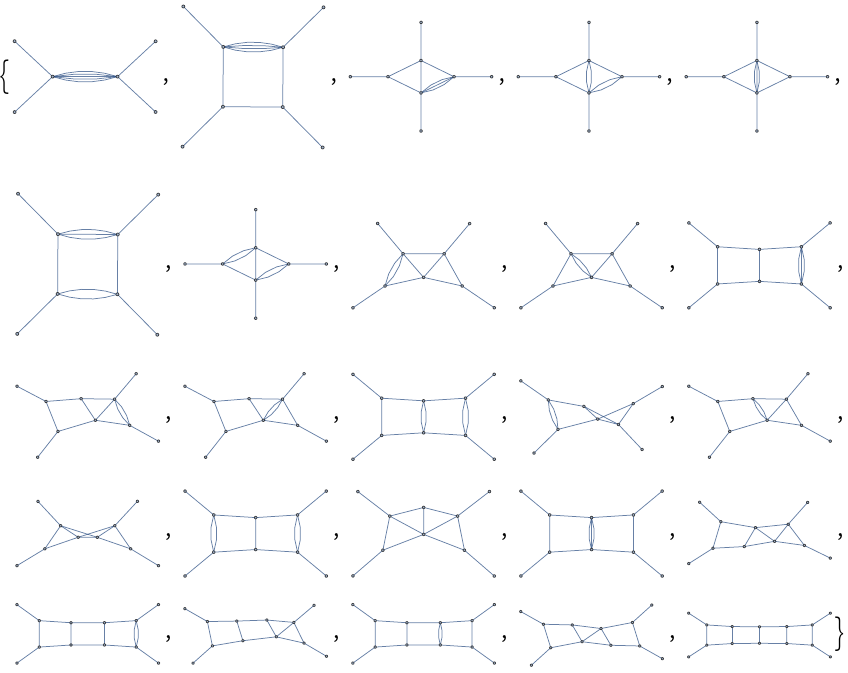}
	\caption{Master integrals without numerators in the four-loop ladder top sector on the $u$ cut.}
	\label{fig:4Ldiags}
\end{figure}

We found 59 MIs $M_{u,i}$ for the ladder top sector on the $u$ cut using \texttt{Kira}~\cite{Maierhofer:2017gsa,Maierhofer:2018gpa,Klappert:2019emp}, see Fig.~\ref{fig:4Ldiags}.
In order to construct the DEQ~\ref{eq:deqMI}, we took derivatives of the MIs with \texttt{Mathematica} package \texttt{LiteRed}~\cite{Lee:2013mka}.
We IBP reduced these derivatives back to the MI basis using \texttt{Kira}~\cite{Maierhofer:2017gsa,Maierhofer:2018gpa,Klappert:2019emp}.
We found a canonical MI basis $M^c_{u,j}$ with a transformation matrix such that $M_{u,i} = T_{ij} M^c_{u,j}$ using \texttt{Mathematica} package \texttt{CANONICA}~\cite{Meyer:2016slj}.
The canonical matrix has a form
\begin{equation}
	A^c_{ij}(x) = \frac{a_{ij}}{x} + \frac{b_{ij}}{1-x} \,,
\end{equation}
where $a_{ij}$ and $b_{ij}$ are rational numbers.
We used \texttt{MultivariateApart}~\cite{Heller:2021qkz} and \texttt{FiniteFlow}~\cite{Peraro:2019svx} to simplify intermediate rational expressions.
When solving the canonical DEQ, we used \texttt{Mathematica} package \texttt{PolyLogTools} to manipulate HPLs up to transcendental weight 5, and our own implementation up to weight 8.
We found a general canonical solution to the canonical DEQ~\ref{eq:deqCan} via a path-ordered exponential
\begin{equation}
	M^c_u(x,\ep) = \mathbb{P} e^{\ep \int A^c(x) dx} M_{u,0}^c(\ep)
\end{equation}
to weight 7.
Consistently with the form of the canonical matrix $A^c$, the solution is expressed in terms of HPLs with letters 0 and 1.
While this completely fixes the kinematic $x$-dependence of the solution, we still need to fix the BCs $M_{u,0}^c(\ep)$.

In order to find further relations between the BCs, we required regularity of the solution near all three singularities and in all six distinct kinematic crossings of the canonical system~\ref{eq:deqCan} to weight 8, following Refs~\cite{Henn:2020lye,Bargiela:2021wuy,Bargiela:2023aiy}.
In this way, we related all BCs to 3 independent ones at weight 7.
Since canonical MI basis is a linear combination of single integrals as depicted in Fig.~\ref{fig:4Ldiags}, the remaining 3 canonical BCs require computing 5 separate integrals.
These integrals are number 1, 6, and 7 in Fig.~\ref{fig:4Ldiags}, together with 2 additional ones resulting from multiplying in ISPs to the integrand of number 6 and 7 in Fig.~\ref{fig:4Ldiags}.
We evaluated these remaining canonical BCs with high precision using \texttt{AMFlow}~\cite{Liu:2022chg}.

Having found the complete solution to the canonical MIs $M^c_u$ on the $u$ cut, we used method~\ref{sec:ansatz2d} to constrain uncut canonical MIs $M^c$ with $\disc_1 M = \cut_{u} M = M_u$ and $\disc_0 M = \cut_{t} M = 0$.
This completely fixes the kinematic $x$-dependence of uncut canonical MIs $M^c$ to weight 8.
The ladder integral in Fig.~\ref{fig:4LMIs} arises from the linear combination $M_{\text{lad}} = T_{{\text{lad}},j} M^c_{j}$.
The resulting expression for $(x-1)M_{\text{lad}}$ is a pure transcendental function.
Since the analytic expression for $M_{\text{lad}}(x=0,\ep)$ is not required to reconstruct the full amplitude using the Integrated Unitarity equation~\ref{eq:IUeq}, we do not provide it here.
The final expression for $\ep^8(x-1)(M_{\text{lad}}(x,\ep) - M_{\text{lad}}(0,\ep))$ is provided in a computer-readable format in ancillary files attached to this draft.

\section{Conclusions}
\label{sec:concl}

There is a plethora of consequences of the Integrated Unitarity equation~\ref{eq:IUeq}.
Their impact is important both for the formal development in mathematical structure of scattering amplitudes as well as for phenomenological applications.
Let us elaborate on some of them.

\paragraph{Multivariate case}

In modern phenomenological applications, multi-loop multi-scale amplitudes are required.
A generalization of eq.~\ref{eq:IUeq} may introduce a dependence on additional external momenta, their masses, as well as on internal masses.
The first two lead to new external channels which can be cut.
The last one fixes the starting points of branch cuts corresponding to multi-particle production in external channels, as well as to poles corresponding to single-particle production.
Anomalous thresholds may also appear~\cite{Correia:2022dcu}.
Moreover, there are Landau singularities originating in specific underlying integral topologies~\cite{Fevola:2023fzn,Fevola:2023kaw,Helmer:2024wax}.
By iteratively applying the univariate Cauchy's formula~\ref{eq:disp} to multiple external variables, and combining with the definition of sequential discontinuities of Ref.~\cite{Bourjaily:2020wvq} similar to the univariate eq.~\ref{eq:disccut}, one arrives at the multivariate generalization of the Integrated Unitarity equation
\begin{widetext}
	\begin{equation}
		\amp(\vec{z}) = \frac{1}{(\tpi)^n} \left( \prod_{m=1}^{n} \int_{x_{\text{thres}}^{(m)}}^{\infty} \frac{dx_m}{x_m-z_m}  \cut_{x_m} \right) \amp(\vec{x})
		+ \dots
		\label{eq:MIUeq}
	\end{equation}
	\begin{figure}[h]
		\centering
		\includegraphics[width=0.8\textwidth]{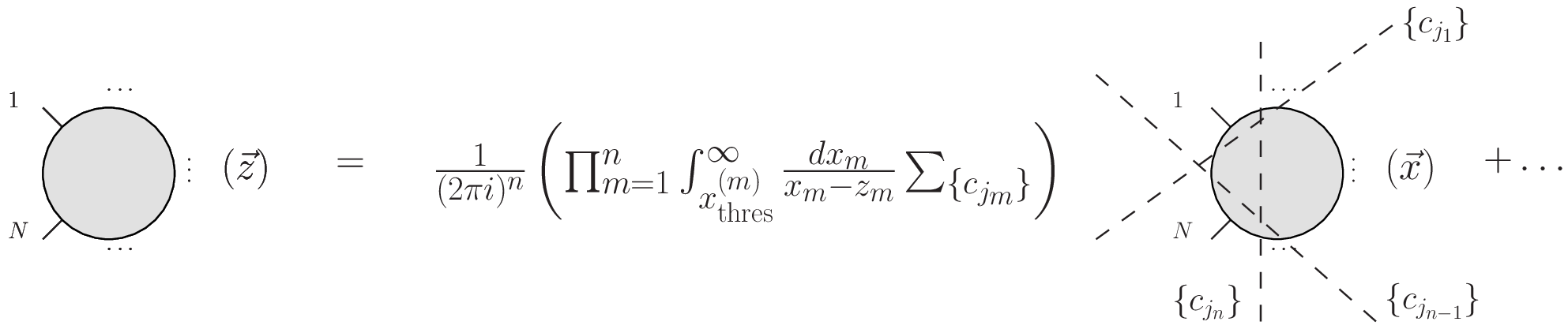}
		\caption{Schematic diagrammatic multivariate generalization of the Integrated Unitarity equation~\ref{eq:MIUeq}.}
		\label{fig:MIU}
	\end{figure}	
\end{widetext}
where $n$ is a number of external variables $\vec{x}$, $x_{\text{thres}}^{(m)}$ corresponds to the branch cut starting point in variable $x_m$, and $+...$ denotes all the contributions from residues and arcs at infinity.
As in the univariate case~\ref{eq:IUeq}, notice a factorization of the amplitude into lower-loop on-shell amplitudes on cuts.
In addition, comparing to eq.~\ref{eq:IUeq}, multiple channels are cut as the same time, which decreases the complexity of the integral reduction even further.
Moreover, it is worth pointing out that some of the sequential discontinuities in overlapping channels may vanish due to Steinmann relations~\cite{Caron-Huot:2016owq,Chicherin:2017dob}.
We also note that multi-scale letters may contain square roots and thus generate discontinuities themselves, which would have to be accounted for systematically.
Another technical complication may arise from the subtleties of multivariate complex analysis.

\paragraph{Recursive approach}

The Generalized Unitarity method~\cite{Britto:2004nc,Anastasiou:2006jv,Ellis:2007br,Giele:2008ve,Ellis:2008ir,Badger:2008cm,Abreu:2017xsl} allows for relating MI coefficient functions to the tree-level on-shell amplitudes.
By using physical principles i.e. causality, which implies analyticity, and unitarity in sec.~\ref{sec:disp}, we have extended the Generalized Unitarity approach and expressed the whole scattering amplitude in terms of full lower-loop on-shell amplitudes.
Due to the recursive nature of the Integrated Unitarity equation~\ref{eq:IUeq}, one can imagine building up an amplitude in a loop-by-loop order starting from tree-level amplitudes.
In practice, one needs to remember that physical helicity amplitudes have purely four-dimensional external momenta, while integrating them along the cut should rely on using full $d$-dimensional states~\cite{Kilgore:2011ta}.
It is also worth pointing out that explicitly using lower-loop results may require integrating iterated integrals with arguments depending on the cut loop momenta.

\paragraph{Kinematic limits}

As discussed in sec.~\ref{sec:disp}, if one is interested only in the transcendental function contribution to the amplitude, one can use only a subset of constraints i.e. $\disc_0$ and $\disc_1$.
The rational function remainder would not be reconstructed in this approach.
This may be used to find limiting behaviour of an amplitude at a fixed logarithmic accuracy to any subleading power.

It is clear that there is a lot of interesting future directions to pursue in order to elucidate all the implications of the Integrated Unitarity equation~\ref{eq:IUeq}.

\section*{Acknowledgements}

We extend our gratitude to T. Gehrmann, S. Abreu, and S. Mizera for valuable comments on this draft.
We thank T. Yang,  K. Sch\"onwald, F. Lange, and T. Peraro for sharing their experience with various packages used in this work.
We are also grateful to V. Sotnikov, G. Falcioni, and R. Marzucca for interesting discussions.
Moreover, we would like to thank the organizers and lecturers of the Higgs Centre School of Theoretical Physics 2023~\cite{Mizera:2023tfe}.
The research of PB was supported by the Swiss National Science Foundation (SNF) under contract 200020-204200 and by the European Research Council (ERC) under the European Union's Horizon 2020 research and innovation programme grant agreement 101019620 (ERC Advanced Grant TOPUP).
Feynman graphs were drawn with \texttt{JaxoDraw}~\cite{Vermaseren:1994je,Binosi:2003yf}.


\bibliographystyle{apsrev4-1}
\bibliography{references}

\end{document}